\documentclass[twocolumn,showpacs,preprintnumbers,amsmath,amssymb,superscriptaddress]{revtex4}
\usepackage{placeins}
\usepackage{graphicx}
\usepackage{amsmath}
\usepackage{times}
\usepackage{color}

\begin{document}
\title{Effects of inhomogeneous activity of players and
noise on cooperation in spatial public goods games}

\author{Jian-Yue Guan}
\thanks{guanjr03@lzu.cn} \affiliation{Institute of Theoretical
Physics, Lanzhou University, Lanzhou Gansu, 730000, China}

\author{Zhi-Xi Wu}
\thanks{wupiao2004@yahoo.com.cn} \affiliation{Institute of
Theoretical Physics, Lanzhou University, Lanzhou Gansu, 730000,
China} \affiliation{Department of Electronic Engineering, City
University of Hong Kong, Kowloon, Hong Kong, China}

\author{Ying-Hai Wang}
\thanks{yhwang@lzu.edu.cn}\affiliation{Institute of Theoretical
Physics, Lanzhou University, Lanzhou Gansu, 730000, China}

\date{Received: date / Revised version: date}
\begin{abstract}
We study the public goods game in the noisy case by considering
the players with inhomogeneous activity teaching on a square
lattice. It is shown that the introduction of the inhomogeneous
activity of teaching of the players can remarkably promote
cooperation. By investigating the effects of noise on cooperative
behavior in detail, we find that the variation of cooperator
density $\rho_C$ with the noise parameter $\kappa$ displays
several different behaviors: $\rho_C$ monotonically increases
(decreases) with $\kappa$; $\rho_C$ firstly increases (decreases)
with $\kappa$ and then it decreases (increases) monotonically
after reaching its maximum (minimum) value, which depends on the
amount of the multiplication factor $r$, on whether the system is
homogeneous or inhomogeneous, and on whether the adopted updating
is synchronous or asynchronous. These results imply that the noise
plays an important and nontrivial role in the evolution of
cooperation.
\end{abstract}
\pacs{89.65.-s, 02.50.Le, 07.05.Tp, 87.23.Kg}

\maketitle

\section{Introduction}
Cooperation plays an important role in the real world, ranging
from biological systems to economic and as well as social systems
\cite{Colman}. Scientists from different fields of natural and
social sciences often resort to evolutionary game theory
\cite{Smith,Gintis} as a common mathematical framework and the
prisoner's dilemma game (PDG) as a metaphor for studying
cooperation between selfish and unrelated individuals
\cite{Gintis}. The PDG, which captures typical pairwise
interactions in many cases, has become the leading paradigm to
explain cooperative behavior\cite{Axelrod,Szabo0}. As an
alternative, the public goods game (PGG), which can be regarded as
a PDG with more than two participators, attracts also much
attention to study the emergence of cooperative behavior
\cite{Kagel}. In a typical PGG played by $N$ agents \cite{Szabo1},
each of them must decide independently and simultaneously whether
or not to invest to a common pool. The collected investment will
be multiplied by a factor $r$, and then is redistributed uniformly
among all players irrespective of their actual contributions. It
was shown that for the values of $1<r<N$, the \emph{ free riders}
or \emph{defectors} (i.e., those who do not invest) will dominate
the whole population \cite{Hauert,Hauert0}.

To provide an escape hatch out of economic stalemate, Hauert
\emph{et al.} have introduced the voluntary participation in such
public goods and found that it results in a substantial and
persistent willingness to cooperate \cite{Hauert1}. Szab\'{o}
\emph{et al.} have studied the voluntary participation in PGG on
square lattice \cite{Szabo1} and found that the introduction of
loners leads to a cyclic dominance of the strategies and promotes
substantial levels of cooperation. a remarkable increase of
cooperation is also observed for those systems where the
inhomogeneous imitation activity is introduced artificially to
characterize the asymmetric and different influence of players to
each other \cite{kim}. In Ref. \cite{Wu}, Wu \emph{et al.} have
studied the PDG with the dynamic preferential rule and found that
cooperation is substantially enhanced due to this simple selection
mechanism. The effect of payoffs and noise on the maintenance of
cooperative behavior has been studied in PDG where two types of
players ($A$ and $B$) with different activity of teaching are
introduced on regular connectivity structures \cite{Szolnoki}. It
was found that the introduction of the inhomogeneous activity of
teaching the players can remarkably promote cooperation
\cite{Szolnoki}.

In the present, we study the effects of noise on the cooperative
behavior by considering the players with the inhomogeneous
activity of teaching in PGG on the square lattice. Our goals are
to investigate $(i)$ whether the inhomogeneous activity of
teaching of the players can promote cooperation in PGG, and $(ii)$
how the noise level affects the cooperative behavior in the PGG.
Our observations suggest that, the cooperative behavior is
remarkably enhanced due to the introduction of the inhomogeneous
activity of teaching. And we also find that the cooperator density
$\rho_C$ changing with the noise level exhibits several different
behaviors, which depends on the values of the multiplication
factor $r$, on whether the system is homogeneous or inhomogeneous,
and also on whether the adopted updating is synchronous or
asynchronous.

\section{The model and simulations}
Motivated by previous research work \cite{Wu,Szolnoki}, we
consider in our model two types of players ($A$ and $B$) who are
distributed randomly on the sites of a square lattice before the
start of the simulation. The fraction of players $A$ and $B$ are
denoted by $\nu$ and ($1-\nu$), respectively. We consider the
compulsory version of a spatial PGG \cite{Szabo1}, i.e., both
types of players are only cooperators (C) or defectors (D). Each
player interacts only with its four nearest neighbors (von Neumann
neighborhood). Thus the size of interaction is $g=5$. The payoff
of each player depends on the number of cooperators $n_{c}$ in the
neighborhood, i.e., the player $x$'s payoff is
\begin{eqnarray}
p_{x}  =  \frac{rn_{c}}{g}-1, & {\rm ~if~} s_{x}=C,\cr p_{x} =
\frac{rn_{c}}{g}-0, & {\rm ~if~} s_{x}=D,
\end{eqnarray}
where $s_{x}$ denotes the strategy of the player $x$, and $r$
represents the multiplication factor on the public good. The
cooperative investments are normalized to unity and $r>1$ must
hold such that groups of cooperators are better off than groups of
defectors.

After each generation, the players try to maximize their
individual payoff by imitating (learning) one of the more
successful neighboring strategies. Following previous studies
\cite{Szabo0,Szabo1,Wu,Szolnoki,Szabo2}, player $x$ will adopt the
randomly chosen neighbor $y$'s strategy $s_{y}$ with a probability
depending on the payoff difference ($p_{x}-p_{y}$) as
\begin{equation}
W(s_{x}\rightarrow
s_{y})=\omega_{xy}\frac{1}{1+exp[(p_{x}-p_{y})/\kappa]},
\end{equation}
where $\kappa$ denotes the amplitude of noise. The pre-factor
$\omega_{xy}$ is given as
\begin{eqnarray}
\omega_{xy}  =  1, & {\rm ~if~} n_{y}=A, \cr \omega_{xy}  =
\omega, & {\rm ~if~} n_{y}=B,
\end{eqnarray}
where the value of $\omega$ ($0<\omega<1$) characterizes the
strength of reduced teaching activity if the site $y$ is occupied
by a player of type $B$ \cite{Szolnoki}.

Simulations were carried out for a population of $N=100\times100$
individuals. We study the key quantity of cooperator density
$\rho_C$ in the steady state. Initially, the two strategies of $C$
and $D$ are randomly distributed among the individuals with equal
probability $1/2$. The above model was simulated with both
synchronous and asynchronous updating. Eventually, the system
reaches a dynamic equilibrium state. The simulation results were
obtained by averaging over the last $10^{4}$ Monte Carlo (MC) time
steps of the total $10^{5}$. Each data point results from an
average over $20$ realizations.

\begin{figure}[h]
\begin{center}
{\includegraphics[width=8cm]{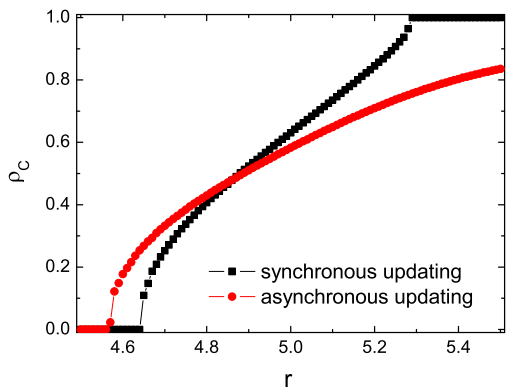}} \caption{(Color online)
The cooperator density $\rho_C$ vs the multiplication factor $r$
when $\nu=1$ by using the synchronous and asynchronous updating.
The noise parameter is $\kappa=0.1$.} \label{fig1}
\end{center}
\end{figure}

\section{Simulation results and analysis}
To facilitate comparison, we first consider the homogeneous system
($\nu=1$), i.e. the system for two types of players with the same
teaching activity. Fig.\ \ref{fig1} shows that the cooperator
density $\rho_C$ varies as the multiplication factor $r$ when
$\nu=1$ by using synchronous and asynchronous updating. We can
find from this figure that, below some threshold values of the
multiplication factor ($r<r_{s}$ for synchronous updating and
$r<r_{as}$ for asynchronous updating), the fraction of cooperators
$\rho_C$ vanishes and the defectors reign; while for the values of
multiplication factor $r>r_{s}$ and $r>r_{as}$, the cooperator
density $\rho_C$ increases with $r$. For large $r$, $\rho_C$ can
reach its maximum value $1$, i.e., the system reaches the
absorbing state of all cooperators, because large values of $r$
are in favor of cooperators in the context of the PGG
\cite{Hauert}. We compare the simulation results for synchronous
and asynchronous updating and find that, when $r$ is smaller than
a certain value $r_c$, $\rho_C$ is larger for asynchronous
updating than that for synchronous updating, but the reverse is
true for $r>r_c$. Here we want to remark that in the process of
our simulation, for synchronous updating, in each time step, all
sites are updated simultaneously through competition with a
randomly chosen neighbor. For asynchronous updating, however, in
each time step, the individuals update their strategies one by one
in a random sequence. It is known that in the context of the PGG,
small values of $r$ favor defectors and large $r$ benefits
cooperators \cite{Hauert,Hauert0}. Comparing the average frequency
of cooperators under the same values of $r$, we find that for
small (large) $r$, defectors (cooperators) live better for
synchronous updating than that for asynchronous updating.

\begin{figure}[h]
\begin{center}
{\includegraphics[width=8cm]{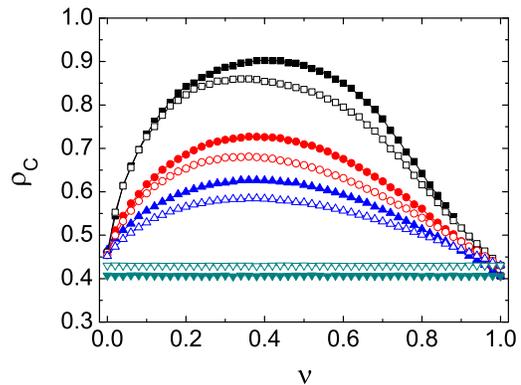}} \caption{(Color online)
The simulation results of cooperator density $\rho_C$ as a
function of $\nu$ for different values of $\omega$ when $r=4.8$
and $\kappa=0.1$ by using synchronous (solid symbols) and
asynchronous (hollow symbols) updating on the square lattice. The
values of $\omega$ from top to bottom are respectively 0.01, 0.1,
0.2, and 1.0.} \label{fig2}
\end{center}
\end{figure}

In Fig.\ \ref{fig2}, we show the variation of cooperator density
$\rho_C$ with the fraction $\nu$ of individuals $A$ for several
different values of $\omega$ when $r=4.8$ and $\kappa=0.1$ by
using synchronous (solid symbols) and asynchronous updating
(hollow symbols). In our model, if $\omega=1$ the system is
equivalent to a homogeneous system (discussed in Fig.\
\ref{fig1}). For $\omega<1$, i.e., the inhomogeneous system, the
cooperator density $\rho_C$ increases monotonously until reaching
the maximum value at $\nu\simeq0.4$ ($\nu\simeq0.33$) for
synchronous (asynchronous) updating, and then it decreases with
$\nu$. And we also find that for the same value of $\nu$, the
smaller the $\omega$ is, the larger the $\rho_C$ will be. This
phenomenon is similar to the results obtained in Ref.
\cite{Szolnoki}. We argue that it is a general phenomenon that the
introduction of reducing activity of teaching can remarkably
enhance the cooperative behavior in the context of whether PDG or
PGG \cite{Note}. From Fig.\ \ref{fig2}, we can also see that the
qualitative results of the cooperative behavior are unchanged for
both synchronous and asynchronous updating. Quantitatively, for
the same value of $\nu$, the cooperator density $\rho_C$ is larger
for synchronous updating than that for asynchronous updating when
$\omega<1$, but the reverse is true in the case of $\omega=1$. It
indicates that, the enhanced level of cooperation due to the
introduction of reducing activity of teaching depends on the
detailed updating (synchronous or asynchronous).

\begin{figure}[h]
\begin{center}
{\includegraphics[width=9cm]{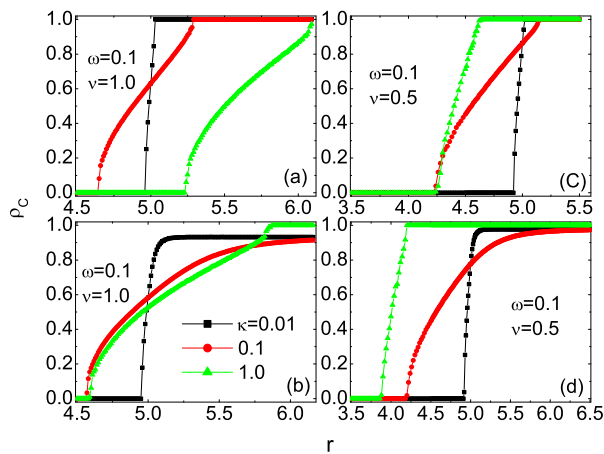}} \caption{(Color online)
The cooperator density $\rho_C$ vs the multiplication factor $r$
in homogeneous (a),(b) and inhomogeneous (c),(d) systems by using
the synchronous (a),(c) and asynchronous (b),(d) updating for
three noise parameters: $\kappa=0.01, 0.1, 1.0$.} \label{fig3}
\end{center}
\end{figure}

\begin{figure}[h]
\begin{center}
{\includegraphics[width=9cm]{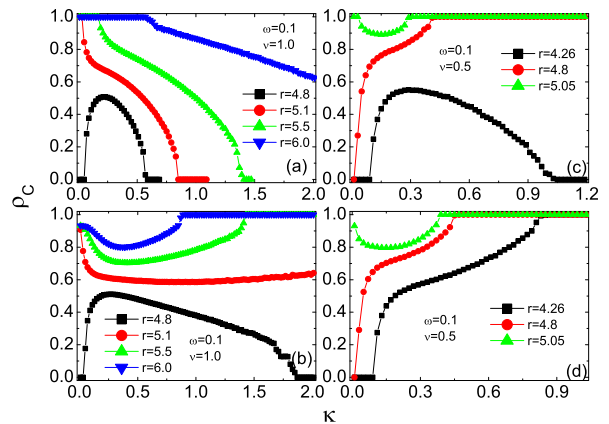}} \caption{(Color online)
The cooperator density $\rho_C$ vs the noise parameter $\kappa$ in
homogeneous (a),(b) and inhomogeneous (c),(d) systems by using the
synchronous (a),(c) and asynchronous (b),(d) updating for
different values of $r$: the detailed values are shown in the
plots.} \label{fig4}
\end{center}
\end{figure}

In our simulations, we find that the level of noise, i.e., the
magnitude of $\kappa$, has  a nontrivial effect on the evolution
of cooperation. We show in Fig.\ \ref{fig3} the cooperator density
$\rho_C$ varying with the multiplication factor $r$ in homogeneous
[Figs.\ \ref{fig3}(a) and (b)] and inhomogeneous [Figs.\
\ref{fig3}(c) and (d)] systems for several different values of
$\kappa$ by using the synchronous [Figs.\ \ref{fig3}(a) and (c)]
and asynchronous [Figs.\ \ref{fig3}(b) and (d)] updating. In the
stationary state $\rho_C$ monotonously increases if $r$ is
increased for the same noise parameter. For the three noise
parameter $\kappa=0.01, 0.1, 1.0$, in homogeneous systems [Fig.\
\ref{fig3}(a) and (b)], when $r=4.8$, $\rho_C$ firstly increases
and then decreases with $\kappa$, while for $r=5.1$, $\rho_C$
monotonously decreases with $\kappa$; for larger values of
$r=6.0$, the cooperator density displays different behaviors with
$\kappa$: $\rho_C$ monotonously decreases with $\kappa$ for
synchronous updating in Fig.\ \ref{fig3}(a); and $\rho_C$ firstly
decreases and then increases with $\kappa$ for asynchronous
updating in Fig.\ \ref{fig3}(b). In inhomogeneous systems (Fig.\
\ref{fig3}(c) and (d)), when $r=4.8$, $\rho_C$ monotonously
increases with $\kappa$; while for $r=5.05$, $\rho_C$ firstly
decreases and then increases with $\kappa$; for smaller values of
$r=4.26$, the cooperator density displays different behaviors with
$\kappa$: $\rho_C$ firstly increases and then decreases with
$\kappa$ for synchronous updating in Fig.\ \ref{fig3}(c); and
$\rho_C$ monotonously increases with $\kappa$ for asynchronous
updating in Fig.\ \ref{fig3}(d). To investigate this in detail, we
plot $\rho_C$ vs $\kappa$ for different values of $r$ by using
synchronous and asynchronous updating in homogeneous and
inhomogeneous systems respectively (see Fig.\ \ref{fig4}).

Before making further discussions about our MC data, we want to
give a brief description of the mean field result for the PGG. In
a well mixed population (i.e., in the mean field case) with
initial number of cooperator $N\rho_C$, each cooperator and
defector will get, respectively, $r\rho_C -1$ and $r\rho_C$
payoff. If a defector (cooperator) changes its strategy to be a
cooperator (defector), then its payoff variation is $r/N-1$ (and
$1-r/N$). For any values of $r$ larger than $N$, one can easily
see that the transformation of defection to cooperation is
preferred by the players, and the whole population will be
dominated by cooperators. In the reverse, for $1<r<N$, defectors
dominate the population \cite{Szabo0,Hauert,Hauert0}. Looking back
on Fig. \ref{fig3}, it is clear that in the low noise limit
$\kappa=0.01$, the MC data are in accordance with the results
predicted by mean field theory \cite{Szabo0,Hauert,Hauert0}, i.e.,
$\rho_C=0$ if $r<5.0$ and $\rho_C=1$ if $r>5.0$ ($N=g=5$ in the
present case).

Figure \ref{fig4} shows that the fraction of cooperators $\rho_C$
varies with the noise parameter $\kappa$ for different values of
$r$ in the two systems [the homogeneous in Figs.\ \ref{fig4}(a)
and \ref{fig4}(b), the inhomogeneous in Figs.\ \ref{fig4}(c) and
\ref{fig4}(d)] by using two updating rules (the synchronous
updating in Fig.\ \ref{fig4} (a) and (c), the asynchronous
updating in Fig.\ \ref{fig4} (b) and (d)). Let us first see the
case in homogeneous systems. In Figs.\ \ref{fig4} (a) and
\ref{fig4}(b), when $r=4.8$, the cooperator density $\rho_C$
firstly increases monotonously until reaching the maximum value,
and then decreases monotonously with $\kappa$, which indicates
that the cooperative behavior can be remarkably enhanced at the
optimal level of noise both for synchronous and asynchronous
updating in homogeneous systems. For larger values of $r$, the
concentration of cooperators $\rho_C$ changing with $\kappa$
displays different behaviors: It decreases monotonously with
$\kappa$ for synchronous updating [see the curves for $r=5.1, 5.5,
6.0$ in Fig.\ \ref{fig4} (a)]; while for asynchronous updating,
when increasing the noise level, one can observe the valleys in
the concentration of cooperators [see the curves for $r=5.1, 5.5,
6.0$ in Fig.\ \ref{fig4} (b)]. That is, a large level of noise
evidently inhibits the persistence of cooperation for synchronous
updating; but for asynchronous updating, the cooperative behavior
can be inhibited mostly at certain value of $\kappa$, which
indicates that the cooperative behavior depends remarkably on the
updating rule. 
In inhomogeneous systems, for synchronous updating [Fig.\
\ref{fig4}(c)], the cooperator density $\rho_C$ varying with
$\kappa$ displays three different behaviors: $\rho_C$ firstly
increases with $\kappa$ and then decreases monotonously after
reaching its maximum value for small value of $r=4.26$; $\rho_C$
increases monotonously with $\kappa$ for moderate value of
$r=4.8$; $\rho_C$ firstly decreases with $\kappa$ and then
increases monotonously after reaching its minimum value for large
value of $r=5.05$. This indicates that the changing behavior of
$\rho_C$ with $\kappa$ are related closely to the multiplication
factor $r$. But for asynchronous updating [Fig.\ \ref{fig4}(d)],
when $r=4.26$, $\rho_C$ increases monotonously with $\kappa$,
which is different from the case for synchronous updating. Once
again, this indicates definitely that the cooperative behavior has
strong dependence on the updating rule.

In addition, for the synchronous updating and the same value of
$r$, we can observe a peak in the concentration of cooperators
when increasing the noise level in homogeneous system [see the
curve for $r=4.8$ in Fig.\ \ref{fig4}(a)], but the cooperator
density $\rho_C$ increases monotonously with the parameter
$\kappa$ in inhomogeneous systems [see the curve for $r=4.8$ in
Fig.\ \ref{fig4}(c)]. It indicates that the evolution of
cooperation also depends on whether the system is homogeneous or
inhomogeneous. From the above behaviors, we can conclude that the
effects of noise on the cooperative behavior remarkably depend on
the values of $r$, on whether the system is homogeneous or
inhomogeneous, and on whether the updating is synchronous or
asynchronous.

\section{Conclusion}
In summary, we have studied the effects of noise on cooperation in
spatial public goods games by considering two types of players
with different activity of teaching. It was shown that the
introduction of the inhomogeneous activity of teaching the players
can remarkably enhance the persistence of cooperation. The
introduction of the inhomogeneous activity of teaching can
partially resolve the dilemma of cooperation and may shed new
light on the evolution of cooperation in the society. By
investigating the effects of noise on the cooperative behavior in
detail, we found that the cooperator density $\rho_C$ varying with
the noise level $\kappa$ displays several different behaviors,
which remarkably depends on the values of $r$, on whether the
system is homogeneous or inhomogeneous, and on whether the
updating is synchronous or asynchronous. Interestingly, when
increasing the level of noise, we observed both peaks and valleys
in the concentration of cooperators for the middle level of noise.
Thus the effect of noise is a correlated factor in game dynamics,
which plays an important role in the evolution of cooperation.

\acknowledgments{This work was supported by the National Natural
Science Foundation of China under Grant No. 10775060.}

\bibliographystyle{h-physrev3}

\begin{thebibliography}{10}
\bibitem{Note}
Recently, we have found that the promotion of cooperation due to
the inhomogeneous activity of the players also holds for other
$2\times 2$ games, for example, the snowdrift game \cite{Szabo0}.
\end{thebibliography}

\end{document}